\documentstyle[aps,epsf,12pt]{revtex}

\def\pmb#1{\setbox0=\hbox{$#1$}%
\kern-.025em\copy0\kern-\wd0
\kern.05em\copy0\kern-\wd0
\kern-.025em\raise.0433em\box0}

\def\beq{\begin{equation}}
\def\eeq{\end{equation}}

\def\Gm0{\Gamma^\mu_0}

\input BoxedEPS
\SetRokickiEPSFSpecial 
\HideDisplacementBoxes 

\begin{document}

\def\footnoterule{\hrule width \hsize}
\def\footstrut{\baselineskip 16pt}

\skip\footins = 14pt 
\footskip     = 20pt 
\footnotesep  = 12pt 

\textwidth=6.5in
\hsize=6.5in
\oddsidemargin=0in
\evensidemargin=0in
\hoffset=0in

\textheight=9.5in
\vsize=9.5in
\topmargin=-.6in
\voffset=-.3in

\baselineskip=18pt plus .5pt

\title{%
\vspace*{12pt}
EVOLUTION OF (WARD--) TAKAHASHI RELATIONS AND HOW I USED THEM
}

\footnotetext[1] {\baselineskip=16pt This work is supported in part by funds
provided by  the U.S.~Department of Energy (D.O.E.) under contract
\#DE-FC02-94ER40818. \hfil MIT-CTP-2669 \hfil  September 1997\break}

\author{R.~Jackiw\footnotemark[1]}

\address{Center for Theoretical Physics\\ Massachusetts Institute of
Technology\\ Cambridge, MA ~02139--4307}

\maketitle

\setcounter{page}{0}
\thispagestyle{empty}

\vskip.5in

\begin{abstract}%
\noindent%

\end{abstract}
The story of (Ward--) Takahashi relations and their impact on physical theory
is reviewed.

\vskip 2in

\centerline{40 Years of Ward--Takahashi Relations, Edmonton, Canada, September
1997}

\maketitle

\newpage


Quantum field theories that describe realistically the fundamental interactions
of elementary particles cannot be solved completely; mostly only approximate
probes are available to us theorists.  Consequently, when it happens that an
exact result can be deduced, we are enormously pleased and cherish the
derivation.  That is why today we are happily commemorating the anniversary of
one such exact observation -- the (Ward--) Takahashi identity -- which has
elucidated the structure of quantum field theories and partially described
their dynamical content, without recourse to any approximation scheme.  This
identity owes its final and general form to our colleague here in Edmonton,
Yasushi Takahashi, and I shall review its evolution and successes over the
years and also describe its impact on my research.

The story begins in 1950, when quantum field theory, more specifically spinor
quantum electrodynamics, was being studied perturbatively and its divergences
were being removed by renormalization.  It soon became apparent that the
multiplicative renormalization constant of the Fermion propagator $S(p)$
should coincide with that of the photon vertex function $\Gamma^\mu (p,q)$,
and that this could be demonstrated if the propagator and vertex were related
in some fashion.  (The arguments of these functions are the four-momenta
carried by the Fermions: a single
$p$ for the propagator; two values, $p$ and $q$, for the vertex function, so
that the photon momentum is $p-q$.)

A result that did the job was found by Ward, who recognized
that a relation between the free expressions $S_0 (p) = \frac{1}{i}
(/ \llap{$p$}-m)^{-1}$ and $\Gamma^\mu_0 (p,q) = \gamma^\mu$, {\it
viz.\/} the formula
$$
\frac{\partial}{\partial p_\mu} S_0^{-1}(p) = i \gamma^\mu = i \Gamma_{0}^{\mu}
(p,p)
$$
also holds, order-by-order in perturbation theory, for the complete Fermion
propagator and photon vertex at coincident arguments\cite{ref:1}.  Ward's
identity
\begin{equation}
\frac{\partial}{\partial p_\mu} S^{-1}(p) = i \Gamma^\mu (p,p)
\label{eq:1}
\end{equation}
directly implies the desired equality of renormalization constants.  Moreover,
immediately and independently, it was realized that Eq.~(\ref{eq:1}) can also
be used to give information about the photon vertex in the forward direction,
for vanishing photon momentum,
$$
\Gamma^\mu (p,q) \,\,\,
\raisebox{-1ex}{\mbox{$\scriptsize \overrightarrow{p-q
\to 0}$}}
\,\,\, -i
\frac{\partial}{\partial p_\mu}S^{-1}(p) 
$$
With this information one can establish, without perturbation theory, exact
low energy theorems for photon absorption, emission and scattering processes.  

The first such threshold theorem was derived for the Compton
amplitude, independently of Ward's result, by Thirring \cite{ref:2}; this was
soon followed by the more refined analyses of Kroll and Ruderman, of Low,
as well as of Gell-Mann and Goldberger, all of whom made explicit use of the
Ward identity\cite{ref:3}.

These successes brought with them new issues.  First, one wondered if the
identity could be established without recourse to the perturbation expansion or
Feynman's diagrams.  Second, one sought a proof for a conjectured\cite{ref:4},
more general relation between the complete propagator and vertex at unequal
Fermion momenta, which had been abstracted from the corresponding free-field
formula.
$$ S^{-1}_0 (p) - S^{-1}_0 (q) = i(p-q)_\mu \Gm0
(p,q) 
$$
Finally, one needed to identify the intrinsic and general property of
quantum field theory that is responsible for the validity of these relations,
which evidently elucidate not only the structure of the theory as seen in its
perturbative expansion, but also determine the infra-red (low-energy) limit of
its dynamics.

The answer to all these questions is found in Takahashi's 1957
paper\cite{ref:5}.  Its content is well summarized by the abstract, which is
here reproduced.

\bigskip
\bigskip
\bigskip
\centerline{\BoxedEPSF{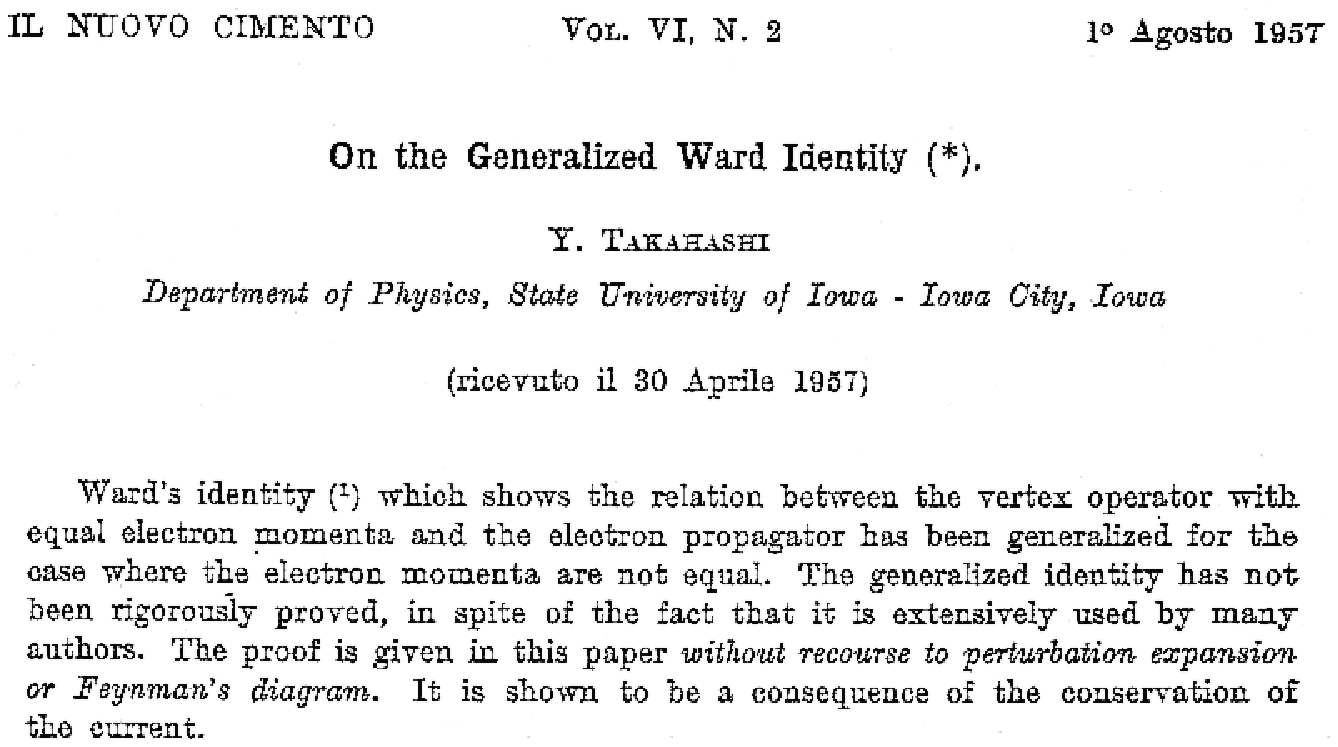}}
\bigskip
\bigskip

The paper provides a non-perturbative proof of the Takahashi
relation 
\begin{equation} 
S^{-1}(p) - S^{-1}(q) = i  (p-q)_\mu \Gamma^\mu (p,q)
\label{eq:2}
\end{equation}
thereby generalizing Ward's identity (\ref{eq:1}) away from the
forward direction $p=q$.  Also Takahashi's derivation
clearly exposes the theoretical underpinnings of the argument.

His key observation is the formula
\begin{equation}
\frac{\partial}{\partial x^\mu} \Bigl(Tj^\mu (x) {\cal O}(0) \Bigr) = [j^0(x), 
{\cal O}(0)] \delta (x^0)+ T \biggl(\frac{\partial}{\partial x^\mu} j^\mu(x) 
{\cal O}(0)\biggr)
\label{eq:3}
\end{equation}
where $T$ signifies time-ordering of the two local operators $j^\mu(x)$ and $
{\cal O}(0)$.  [In deriving (\ref{eq:2}), Takahashi let $j^\mu$ be the
conserved electromagnetic current; he used two local operators instead of the
single $ {\cal O}$ --- two Fermi fields at different space-time points; and he
worked in the Fourier transformed momentum space; but the essential ideas are
already contained in (\ref{eq:3}).]  The first term on the right, involving the
$\delta$-function in time $(x^0)$, arises from differentiating the
discontinuities of the
$T$ product and renders the coefficient commutator to be at equal times. The
second term contains the divergence of the current operator, which vanishes if
$j^\mu$ is a symmetry current, as in Takahashi's application.  The equal-time
commutator may be deduced with canonical commutation relations, when an
explicit formula for $j^0$ is available --- that is, one can use the
correspondence principle to infer commutators from Poisson brackets.  Another
argument is based on the observation that the spatial volume integral of $j^0$,
$Q = \int dV j^0$, generates by commutation some definite transformation on all
dynamical variables.  Indeed if $j^\mu$ corresponds to a conserved symmetry
current, $Q$ is the time-independent symmetry generator.  If that
transformation is explicitly known for the quantity ${\cal O}$ --- call it
$\Delta {\cal O}$ --- than one has $i[Q, {\cal O}(0)] = \Delta {\cal O}(0)$ and
it is plausible to suppose that 
\begin{equation}
i[j^0(x), {\cal O}(0)] \delta (x^0) = \Delta {\cal O}(0) \delta (x^0) \delta
(V)
\label{eq:4}
\end{equation}
Even if no symmetry is present so the current is not conserved, formula
(\ref{eq:3}) can still provide information if an
argument can be given for accepting (\ref{eq:4}), and if the divergence of the
current,
$\partial_\mu j^\mu$, is an operator with known properties (in which case one
says that $j^\mu$ is {\it partially conserved\/}).  Takahashi's analysis was
generalized to the case of
$T$ products with several currents by Kazes
\cite{ref:6}.  

The derivation establishes that the foundations of the
(Ward--) Takahashi identity and its generalizations rest on the broad basis of
equal-time commutators, transformations and symmetries, (rather than on the
more restrictive conditions of gauge-invariance\cite{ref:7}), and demonstrates
that the dynamical consequences --- the low-energy theorems
\cite{ref:2,ref:3} --- are truly non-perturbative.

The roles of these identities in quantum
electrodynamics were played out by the 1960's.  But just then quantum field
theory research advanced in a new direction, which became accessible thanks to
the information contained in the generalized Takahashi relations.  

At that time, physicists possessed neither a field theoretical
model for fundamental interactions (other than electromagnetism) nor could they
solve any proposed candidate theory to see whether it is viable.  But it was
appreciated that matrix elements of various vector and
axial-vector currents, corresponding to various internal groups, govern
experimentally accessible processes and carry important information about
fundamental interactions.  Absence of a
realistic theoretical model and of an effective calculational method stymied
progress: the values of these matrix elements could not be 
determined {\it a priori\/}.

The breakthrough came when Gell-Mann postulated his ``current algebra," giving
explicit form to the commutators (\ref{eq:4}).  For $j^0$ he took the time
components of the physically interesting currents; for ${\cal O}$, the full
four-vector or the axial four-vector current; and for $\Delta {\cal O}$, the
expected transformation that follows from the relevant group structure: {\it
i.e.\/}
$\Delta {\cal O}$ is the infinitesimal group rotation of ${\cal O}$.  Also
a definite form for the divergences of these currents was assumed: conserved
for the approximate vector symmetries; partially conserved or spontaneously
broken for the axial-vector currents\cite{ref:8}.  With this information, the
right side of the Takahashi identity (\ref{eq:3}) is explicitly known and
serves to constrain the left side -- {\it i.e.\/} the matrix element of the
physically interesting current.

Thereupon followed an explosion of activity that resulted in an understanding
of spontaneous symmetry breaking and of the Nambu--Goldstone as well as the
Anderson--Higgs phenomena; in low energy theorems, principally for
pseudoscalar mesons that couple to axial-vector currents; in sum rules for
scattering processes; and in a variety of other calculations that mostly
agreed well with experimental observation\cite{ref:9}.  These successes also
provided an enthusiastic vote of confidence for operator quantum field theory,
since current algebra and the Takahashi identities do not fit into any other
frame.

This was the time that I began physics research.  As a graduate student,
I had been frustrated by the absence of a quantum field theoretic
description for non-electromagnetic fundamental processes, so I was delighted
to learn about Gell-Mann's suggestion and the subsequent progress in current
algebra.  Earlier, while studying quantum mechanics, I was very much impressed
that by postulating a largely model-independent equal-time commutator algebra
for position operators of the electron, $i \Bigl[\frac{d}{dt}
r^i, r^j
\Bigr] =
\frac{1}{m} \delta^{ij}$ ($m=$ electron mass), one could derive the
Thomas--Reiche--Kuhn sum rule for transition probabilities, without solving
any dynamical equations or even adopting any particular interaction model.  I
recognized in current algebra a quantum field theoretic analog of the
successful quantum mechanical program for sum rules\cite{ref:10} and decided
to enter into that research.  

But the
subject had reached maturity, with little room for new discoveries. 
However, with collaborators, we managed to find yet another set of low energy
theorems: threshold relations for gravitons
\cite{ref:11} that are completely analogous to the ones for
photons\cite{ref:2,ref:3}.  Since graviton
emission is not yet experimentally accessible the results are academic; but
they are interesting in that novel Takahashi relations, based on Poincar\'e
symmetry and energy-momentum conservation, are used to analyze processes for
which no quantum theory exists: quantum gravity has not yet been
constructed!  This demonstrates vividly the power of these relations in
making physical predictions.

The success of current algebra was marred by uncertainty over the actual
existence of a  field theoretical model in which the
postulated relations are realized and the theorems about dynamics are true. 
Indeed a potential obstacle was recognized from the beginning of the (Ward--)
Takahashi/current algebra program.  It was known that local equal time
commutators frequently differ from the respective Poisson brackets; the
correspondence principle fails.  Specifically when in (\ref{eq:4}) ${\cal O}$
is the spatial component of the current $j^\mu$, there are further
contributions, proportional to spatial derivatives of $\delta$-functions. 
These terms, called {\it Schwinger terms\/}, were exhibited by Takahashi's
compatriots, Goto and Imamura\cite{ref:12}, but in fact they were discovered
by Jordan in the 1930's\cite{ref:13}.  Lack of {\it a priori\/} information
about the Schwinger terms prevents evaluation of (\ref{eq:4}) and of the
right side in (\ref{eq:3}).  However, it was also noted that physical
amplitudes differ from time-ordered products by further local terms, called
{\it seagulls\/}, and in explicit examples, like the Compton scattering
amplitude for scalar electrodynamics, one verified that the divergence of the
seagulls cancels exactly the Schwinger terms, leaving the ``naive''
result, obtained by ignoring the problem altogether.  Taking an optimistic
position, Feynman conjectured that this cancellation is generally valid, and
that one should work with unmodified Takahashi identities.

While such pragmatism allowed calculations to proceed fearlessly, it did not
satisfy me, especially since there remained discrepancies between
some current algebra predictions and experimental facts.  At CERN, where
I came as a visiting researcher, Bell emphasized especially that the current
algebra analysis of neutral pion decay, performed by his colleages Sutherland
and Veltman, leads to the unacceptable conclusion that a massless pion does not
decay into two photons.  But a non-vanishing decay time had been measured and
its magnitude could not be attributed to the small pion mass.  Bell stressed
that the subject of current algebra must not be closed without resolving this
puzzle.  

I decided to elucidate this question, but at first made little progress
because the problematic conclusion presented itself in very immediate
form, being a straightforward consequence of Takahashi identities.  One could
consider, in the Heisenberg picture, the vacuum-two photon matrix element of
the axial-vector current
$j^\mu_5$, to which the pion couples,
$$
T^\mu (z)  = \langle \gamma \gamma | j^\mu_5 (z) | 0 \rangle
$$
$T^\mu$ must be invariant against gauge transformations on the two photons. 
Also in the limit of massless pions, $j^\mu_5$ was taken to be
conserved thanks to chiral symmetry, which was believed to hold in that
limit.  Consequently, $T^\mu$ should be transverse.  Alternatively, one could
work in the electromagnetic interaction picture and consider the vacuum
amplitude of three currents: two electromagnetic currents to which the two
photons couple and $j^\mu_5$
$$
T^{\alpha \beta \mu } (x,y,z) = e^2 \langle 0 |T j^\alpha(x) j^\beta(y)
j^\mu_5 (z) | 0 \rangle 
$$
(On photon mass shell $T^{\alpha \beta \mu }$coincides with $T^\mu$.)  Now the
conservation of all three currents (which satisfy free field equations in the
interaction picture) together with the current algebra assumption of vanishing
commutators leads to transversality of $T^{\alpha
\beta \mu }$ in all its three indices.  Either of these two generalized
Takahashi identities prohibits massless pion decay, and it was difficult to
see how the conclusion could be evaded.

A civilized activity at CERN consists of taking an afternoon drink at the
cafeteria.  Bell and I frequently went there, together with other people who
joined us for discussions.  On one occasion, Steinberger was at the table and
asked about our current interests.  When we described to him our puzzlement
about
$\pi^0
\to 2\gamma$, he expressed amazement that theorists should still be pursuing a
process that he, an experimentalist, had calculated twenty years
earlier, finding excellent agreement with experiment\cite{ref:14}. 

There at the cafeteria table, Bell and I realized that Steinberger's
calculation would be identical to the one performed in the dynamical framework
of the $\sigma$-model, which was constructed to realize current algebra
explicitly.  We reasoned that within the $\sigma$-model, we could satisfy the
current algebraic assumptions of Sutherland--Veltman and also obtain good
experimental agreement in view of Steinberger's result, thereby resolving the
$\pi^0 \to 2\gamma$ puzzle.

Guided by Steinberger's paper (at that time, we were not familiar with similar
 work by Takahashi's compatriots Fukuda and Miyamoto
and only dimly aware of Schwinger's position-space  reanalysis of the
Fukuda--Miyamoto--Steinberger momentum-space calculation\cite{ref:15}) we
quickly established that the lowest order amplitude describing correlations
between the three currents appearing in the problem is given to lowest (one-loop) perturbative order by the now famous
triangle graph and that the explicit result does not obey the expected
Takahashi identities: the single index, vacuum-two photon matrix element of
$j^\mu_5$, 
$T^\mu(z)$, is not transverse when photon gauge invariance is enforced; the
three-index, three-current amplitude $T^{\alpha \beta \mu} (x,y,z)$ cannot be
transverse in all three indices --- transversality is possible in at most
two.  Since the Sutherland--Veltman argument relies on transverality of these
amplitudes, the undesirable conclusion does not hold\cite{ref:16}.

From this calculation, Takahashi identities took on a new life, because in
modified form, they summarize the violation of the correspondence principle by
recording the ``anomalies'' that violate the naive, expected formulas.  In the
case of the vacuum-two photon matrix element of $j^\mu_5$, $T^\mu(z)$, the
anomaly resides in the divergence of the axial vector current: rather than
being conserved it satisfies 
$$
\partial_\mu j^\mu_5 \propto \!\!\!\! \phantom{F}^* \! F^{\mu\nu}
F_{\mu\nu} \ ,
$$ 
where $F_{\mu\nu}$ is the electromagnetic field strength and 
$\hskip-8pt\phantom{F}^* \! F^{\mu\nu}$ is its dual\cite{ref:17}.  On the
other hand, when considering the three current vacuum correlator $T^{\alpha
\beta
\mu } (x,y,z)$, all currents are conserved --- they are free in the interaction
picture --- but the anomaly resides in the commutator algebra: rather than
vanishing commutators, one has an unexpected Schwinger term
$$
[ j^0(x), j^0_5(y) ] \delta(x^0-y^0) \propto B^i (y)
\delta(x^0-y^0) \partial_i \delta^3({\bf x-y})
$$
which does not cancel with the corresponding seagull, leading to the
anomalous Takahashi relation  (here $B^i$ is the magnetic field). 
Consequently Feynman's conjecture is violated\cite{ref:18}.    Anomalous
divergences and anomalous commutators are different sides of the same coin,
and they are unified by giving an altered form to  the Takahashi relation.

After the above discovery resolved the pion decay puzzle, there followed a
search for departures from the correspondence principle in other (Ward--)
Takahashi relations.  These were found in the so-called trace identities,
which involve the trace of the energy-momentum tensor and result from
scale/conformal invariance.  Just as in the previous example of chiral
symmetry, scale/conformal symmetry in quantum field theory also acquires an
anomaly\cite{ref:19}, and the modified Takahashi identities for this case
were recognized as expressions of the renormalization group --- an analysis of
scaling violation in quantum field theory that had been carried out many years
before\cite{ref:20}.

Understanding the departures from naive, canonical reasoning came more or
less at the same time as the construction of the ``standard'' quantum field
theoretic Yang--Mills model of all fundamental interactions (excluding
gravity). Within this model the (Ward--) Takahashi identities
--- for the non-Abelian context they became known as Slavnov--Taylor identities
--- again aided in the renormalization procedure, which is successful only if
the identities are free of anomalies.  This could be achieved provided
quarks and leptons are precisely balanced in number and charge.  Just
such a balance is observed, which is striking evidence that not only quantum
field theorists, but also Nature knows about the subtleties of the Takahashi
identities.

On the other hand, identities that play no role in renormalization do contain
contributions that violate the correspondence principle and lead to a variety
of physical effects in the standard model\cite{ref:21}.  For example unwanted
symmetries, like scale invariance and the pion-decay forbidding chiral
invariance, disappear from the quantum field theory, though they are present
in the classical theory, before quantization.  Another instance is an
unexpected and unwanted source of $CP$ violation, caused by quantum
tunneling.  Perhaps most intriguing is 't Hooft's demonstration that the
anomaly in the Takahashi identity for the baryon number current allows baryons
to decay, but fortunately at a sufficiently slow rate so that the threat to our
physical world is negligible\cite{ref:22}.

In addition to the physics that has emerged from the study of (Ward--)
Takahashi relations, there also arose a very fruitful exchange of ideas with
mathematics, which is still flourishing.  This happened
because it was noticed that the entities that characterize the modifications to
the Takahashi identities, be they  anomalous divergences of currents or
anomalous commutators, possess mathematical significance for
geometry and topology \cite{ref:21}.  The subsequent mathematics-physics collaboration is
most vividly seen these days in the string program; if it lives up to
the promises that its supporters proclaim, it will surely provide a future
arena for new roles to be played by (Ward--) Takahashi relations.

\vtop to 0pt{I wish Yasushi Takahashi a long life, so he can witness and enjoy
the further development of his identities, which evolved from a simple
observation in a corner of physics into a concept that is relevant to much of
the formalism that we use in describing Nature at its fundamental workings.
\vss}

\end{document}